\documentclass[twocolumn,dvipsnames]{aastex62}
\pdfoutput=1 
\usepackage{amsmath,amstext}
\usepackage{apjfonts}
\usepackage[figure,figure*]{hypcap}
\usepackage{ulem}
\usepackage{xspace}
\usepackage{lineno}

\usepackage{xfrac}
\usepackage{graphicx}
\usepackage{natbib}
\usepackage{amssymb}
\usepackage{amsmath}
\usepackage{array,booktabs}
\usepackage{mathptmx}
\usepackage{booktabs}
\usepackage{hyperref}
\usepackage{float}
\usepackage{todonotes}
\usepackage{multirow}

\DeclareSymbolFont{cmletters}{OML}{cmm}{m}{it}
\DeclareMathSymbol{v}{\mathalpha}{cmletters}{"76}

\newcommand{\vej}{\,{v_{\rm ej}}}
\newcommand{\s}{\,{{\rm s}}}
\newcommand{\ms}{\,{{\rm ms}}}

\newcommand{\tmad}{\,{T_{\rm MAD}}}
\newcommand{\thmns}{\,{T_{\rm HMNS}}}
\newcommand{\Mtot}{\,{M_{\rm tot}}}
\newcommand{\mbh}{\,{M_{\rm BH}}}
\newcommand{\mns}{\,{M_{\rm NS}}}
\newcommand{\mdisk}{\,{M_{d}}}

\newcommand{\Eisog}{\,{E_{{\rm iso},\gamma}}}

\newcommand{\msun}{\,{M_{\odot}}}
\newcommand{\erg}{\,{{\rm erg}}}

\newcommand{\cbGRB}{{{\rm cbGRB}}\xspace}
\newcommand{\cbGRBs}{{{\rm cbGRBs}}\xspace}
\newcommand{\lbGRB}{{{\rm lbGRB}}\xspace}
\newcommand{\lbGRBs}{{{\rm lbGRBs}}\xspace}
\newcommand{\sbGRB}{{{\rm sbGRB}}\xspace}
\newcommand{\sbGRBs}{{{\rm sbGRBs}}\xspace}

\shorttitle{A Unified Picture for the Origin of Compact Binary GRBs}
\shortauthors{Gottlieb et al.}

\begin{document}

\title{A Unified Picture of Short and Long Gamma-ray Bursts from Compact Binary Mergers}

    \author[0000-0003-3115-2456]{Ore Gottlieb}
	\email{ogottlieb@flatironinstitute.org}
	\affiliation{Center for Computational Astrophysics, Flatiron Institute, New York, NY 10010, USA}
    \affiliation{Center for Interdisciplinary Exploration \& Research in Astrophysics (CIERA), Physics \& Astronomy, Northwestern University, Evanston, IL 60202, USA}
    \affiliation{Department of Physics and Columbia Astrophysics Laboratory, Columbia University, Pupin Hall, New York, NY 10027, USA}

    \author[0000-0002-3635-5677]{Brian D. Metzger}
    \affiliation{Department of Physics and Columbia Astrophysics Laboratory, Columbia University, Pupin Hall, New York, NY 10027, USA}
    \affiliation{Center for Computational Astrophysics, Flatiron Institute, New York, NY 10010, USA}

    \author[0000-0001-9185-5044]{Eliot Quataert}
    \affiliation{Department of Astrophysical Sciences, Princeton University, Princeton, NJ 08544, USA}

    \author[0009-0005-2478-7631]{Danat Issa}
	\affiliation{Center for Interdisciplinary Exploration \& Research in Astrophysics (CIERA), Physics \& Astronomy, Northwestern University, Evanston, IL 60202, USA}

    \author[0000-0002-1025-8318]{Tia Martineau}
	\affiliation{Department of Physics and Astronomy, University of New Hampshire, 9 Library Way, Durham, NH 03824, USA}

    \author[0000-0003-4617-4738]{Francois Foucart}
	\affiliation{Department of Physics and Astronomy, University of New Hampshire, 9 Library Way, Durham, NH 03824, USA}

    \author[0000-0002-0050-1783]{Matthew D. Duez}
    \affiliation{Department of Physics \& Astronomy, Washington State University, Pullman, Washington 99164, USA}

    \author[0000-0001-5392-7342]{Lawrence E. Kidder}
    \affiliation{Cornell Center for Astrophysics and Planetary Science, Cornell University, Ithaca, New York, 14853, USA}

    \author[0000-0001-9288-519X]{Harald P. Pfeiffer}
    \affiliation{Max Planck Institute for Gravitational Physics (Albert Einstein Institute), D-14467 Potsdam, Germany}

    \author[0000-0001-6656-9134]{Mark A. Scheel}
    \affiliation{TAPIR, Walter Burke Institute for Theoretical Physics, MC 350-17, California Institute of Technology, Pasadena, California 91125, USA}

\begin{abstract}

The recent detections of the $\sim10$-s long $\gamma$-ray bursts (GRBs) 211211A and 230307A followed by softer temporally extended emission (EE) and kilonovae, point to a new GRB class. Using state-of-the-art first-principles simulations, we introduce a unifying theoretical framework that connects binary neutron star (BNS) and black hole--NS (BH--NS) merger populations with the fundamental physics governing compact-binary GRBs (cbGRBs). For binaries with large total masses $M_{\rm tot}\gtrsim2.8\,M_\odot$, the compact remnant created by the merger promptly collapses into a BH, surrounded by an accretion disk. The duration of the pre-magnetically arrested disk (MAD) phase sets the duration of the roughly constant power cbGRB and could be influenced by the disk mass, $M_d$. We show that massive disks ($M_d\gtrsim0.1\,M_\odot$), which form for large binary mass ratio $q\gtrsim1.2$ in BNS or $q\lesssim3$ in BH--NS mergers, inevitably produce 211211A-like long cbGRBs. Once the disk becomes MAD, the jet power drops with the mass accretion rate as $\dot{M}\sim t^{-2}$, establishing the EE decay. Two scenarios are plausible for short cbGRBs. They can be powered by BHs with less massive disks, which form for other $q$ values. Alternatively, for binaries with $M_{\rm tot}\lesssim2.8\,M_\odot$, mergers should go through a hypermassive NS (HMNS) phase, as inferred for GW170817. Magnetized outflows from such HMNSs, which typically live for $\lesssim1\,{\rm s}$, offer an alternative progenitor for short cbGRBs. The first scenario is challenged by the bimodal GRB duration distribution and the fact that the Galactic BNS population peaks at sufficiently low masses that most mergers should go through a HMNS phase.

\end{abstract}

\section{Introduction}\label{sec:introduction}

Gamma-ray bursts (GRBs) can originate from at least two distinct astrophysical systems: the collapse of massive rotating stars (``collapsars''; \citealt{Woosley1993,MacFadyen1999}) and compact binary mergers \citep{Eichler1989,Paczynski1991}. These two event classes are commonly associated with long GRBs (lGRBs) and short GRBs (sGRBs), respectively. Their durations follow log-normal distributions, with mean values of $ \sim 30\,\s $ for lGRBs and $ \sim 0.5\,\s $ for sGRBs \citep{Kouveliotou1993,McBreen1994}. The overlap of the two distributions poses a challenge to a clear distinction between the classes \citep{Bromberg2013}, particularly for bursts lasting between $ \sim 1\,\s $ and $ \sim 30\,\s $ \citep{Nakar2007}. A more accurate burst classification can be obtained when the GRB is followed by optical emission from the astrophysical site: supernova Ic-BL \citep{Galama1998,Hjorth2003} or kilonova from a compact object merger \citep{Li1998,Metzger2010a,Tanvir2013}. Being the most luminous events in the sky, GRBs are detected out to large distances, and in part because of their bright synchrotron afterglows, are infrequently accompanied by detectable thermal optical counterparts.

The recent detection of optical/infrared kilonova signals following two $ \sim 10\,\s $-long bursts in GRB 211211A \citep{Rastinejad2022,Troja2022,Yang2022,Zhang2022} and GRB 230307A \citep{Levan2023b,Sun2023,Yang2023} has reignited interest in the origin of long-duration GRBs that are not associated with collapsars \citep[see also][]{Gal-Yam2006,Della-Valle2006,Bromberg2013,LU2022,Levan2023}, but likely originating from compact binary mergers (\cbGRBs). Such long durations would at least naively be unexpected in binary mergers insofar that the accretion timescales responsible for the jet launching are expected to be of order of seconds (e.g., \citealt{Narayan+92}). The long-duration cbGRB (\lbGRBs) events may constitute a third type of GRB population. Indeed, a closer examination of the GRB duration distribution reveals that it is best fit with three log-normal distributions \citep{Horvath2016,Tarnopolski2016}. These distributions potentially correspond to three distinct populations: (i) collapsar lGRBs with $ T_{90} \gtrsim 30\,\s $, (ii) short-duration \cbGRBs (\sbGRBs) from binary mergers with $ T_{90} \lesssim 1\,\s $, and (iii) \lbGRBs 211211A and GRB 230307A-like events from binary mergers, lasting $ T_{90} \sim 10\,\s $. Below we adhere to the conventional assumption that \sbGRBs are more common than \lbGRBs \citep{Yin2023}. However, we note that three log-normal distribution fits suggest otherwise \citep{Horvath2016}, so we do not consider the rates to be a stringent constraint.

It is tempting to associate the two \cbGRB classes with the two types of compact binary mergers: black hole (BH) and neutron star (NS), and binary NS (BNS) systems. Based on the two BH--NS mergers detected during the LVK O3b run, the BH--NS merger rate was constrained to be $\mathcal{R}_{\rm BHNS} = 45^{+75}_{-33}~{\rm Gpc^{-3}yr^{-1}}$ if these two events are representative of the entire population, versus $\mathcal{R}_{\rm BHNS} = 130^{+112}_{-69}~{\rm Gpc^{-3}yr^{-1}}$ for a broader BH--NS population \citep{Abbott2020}. In comparison, the rate of BNS mergers was found to be $\mathcal{R}_{\rm BNS} = 320^{+490}_{-240}~{\rm Gpc^{-3}yr^{-1}}$ \citep{Abbott2021b}. Therefore, if the two detected BH--NS events are representative, BH--NS mergers are likely to be significantly rarer than BNS mergers, similar to the scarcity of \lbGRBs compared to \sbGRBs. In the case of a broader BH--NS population, other merger properties such as larger mass ratios, significant spin-orbital misalignment, and low BH spins need to be considered \citep{Belczynski2008}, all of which would result in less massive disks and the associated challenges in launching a relativistic jet \citep[e.g.,][]{Kyutoku2015}. Regardless of the BH--NS merger rate, the fraction of this population that yields electromagnetic emission is thus likely to be negligible compared to BNS mergers \citep{Fragione2021,Sarin2022,Biscoveanu2023}.

The main \cbGRB emission phase is often accompanied by additional light curve components. For example, in \lbGRB 211211A, the variable hard burst that lasted $ \sim 10\,\s $ was preceded by an oscillating precursor flare \citep{Xiao2022}, and followed by a smoother and softer $\gamma$/X-ray emission for $ \sim 100\,\s $ \citep{Gompertz2023}, referred to as the "extended emission" \citep[EE;][]{Norris2006,Perley2009}. The prolonged EE, which accompanies the main signal in $ \sim 25\% - 75\% $ of \cbGRBs \citep[][]{Norris2008,Norris2010,Kisaka2017}, is generally characterized by two components: an initial roughly flat ``hump'' \citep{Mangano2007,Perley2009}, followed by a power-law decay $ \sim t^{-2} $ \citep[][]{Giblin2002,Kaneko2015,Lien2016}. Any \cbGRB model linked to the underlying physics of binary mergers must therefore explain the entire emission signal, including precursor flares and EE phases.

One of the main uncertainties in \cbGRB models is the origin of the relativistic jets. They can be generated either through electromagnetic processes from a rotating BH \citep[][hereafter BZ]{Blandford1977} or a magnetized NS \citep[e.g.,][]{Goldreich1969,Usov1992,Duncan1992,Thompson1994,Metzger2011}, or hydrodynamically by the pair plasma produced by annihilation of neutrinos emitted from the accretion disk along the polar accretion funnel \citep[e.g.][]{Eichler1989,Paczynski1990,Woosley1993,MacFadyen1999}. Despite significant progress following the multi-messenger and multi-wavelength event of GW170817 \citep[see][for reviews]{Nakar2020,Margutti2021}, and numerous advanced first-principles simulations of BNS and BH--NS mergers \citep[e.g.][]{Rosswog2003,Shibata2006c,Rezzolla2011,Etienne2012,Hotokezaka2013b,Nagakura2014,Kiuchi2015,Kiuchi2015b,Paschalidis2015,Kawamura2016,Ruiz2016,Ruiz2018,Ruiz2020,Ciolfi2017,Ciolfi2019,Ciolfi2020,Mosta2020,Hayashi2022,Hayashi2023,Sun2022,AguileraMiret2023,Combi2023,Kiuchi2023,Gottlieb2023b}, the connection between the central engine and the aforementioned observed characteristics remains poorly understood.

In this paper, we review recent first-principles simulations, and how they constrain the origins of the different types and phases of \cbGRB light curves. In particular, we present a framework for connecting the binary merger population with the entire spectrum of \cbGRB observations, which provides a first-principles explanation for the origin of the constant-power prompt emission and decaying EE. The paper is structured as follows. In \S\ref{sec:lgrb} we argue that while lGRB jets are powered by magnetically arrested disks (MADs), BH-powered \cbGRB jets are generated before the disk enters a MAD state. In \S\ref{sec:massive} we show that the formation of a massive disk ($ \mdisk \approx 0.1\,\msun $) around the post-merger BH inevitably powers \lbGRBs such as GRB 211211A. In \S\ref{sec:origin} we present two self-consistent models as the origin of \sbGRBs: prompt-collapse BHs forming low-mass disks and hypermassive NSs (HMNSs); we describe why we favor these two scenarios over alternatives, such as delayed-collapse BHs, supramassive NSs (SMNSs), white dwarf (WD) mergers/accretion induced collapse (AIC), and neutrino-driven jets. In \S\ref{sec:comparison} we discuss the origin of the precursor and EE of \cbGRBs, compare the models with observables, and deduce that \sbGRBs are likely powered by HMNSs, whereas \lbGRBs are powered by BHs with massive disks. We summarize and conclude in \S\ref{sec:summary}.

\section{Collapsar GRBs vs. CBGRBs: To be MAD or not to be MAD}\label{sec:lgrb}

Long GRBs and \cbGRBs take place in very different astrophysical environments, leading to distinct conditions for their occurrence and potentially differing central engines that drive these events. A recent study by \citet{Gottlieb2023a} demonstrated that lGRB jets are launched from BHs once the accretion disk becomes MAD. The reason for this is that a successful jet launching requires the Alfv\'{e}n velocity to surpass the free-fall velocity of the inflowing gas, allowing magnetohydrodynamic waves to escape from the BH ergosphere and form the emerging jet \citep{Komissarov2009}. In other words, a sufficiently powerful magnetic flux empowers a BH to launch jets in defiance of the inward motion of the surrounding stellar envelope. Numerical simulations \citep{Gottlieb2022a} have confirmed that this process is sustained once the disk becomes MAD, occurring when the dimensionless magnetic flux on the BH reaches a threshold of $ \phi \equiv \Phi(\dot{M}r_g^2c)^{-1/2} \approx 50 $, where $ r_g $ is the BH gravitational radius, $ \Phi $ is the dimensional magnetic flux, and $ \dot{M} $ is the mass accretion rate \citep[e.g.,][]{Tchekhovskoy2015}. The BZ-jet power is determined by \citep{Blandford1977,Tchekhovskoy2011}:
\begin{equation}\label{eq:BZ}
    P_j \sim \frac{c}{r_H^2}\Phi^2f(a)\,,
\end{equation}
where $ r_H $ is the radius of the BH horizon, and $ f(a) $ is the functional dependency on the BH spin. This relation can also be expressed in terms of the dimensionless magnetic flux $ \phi $:
\begin{equation}\label{eq:Pj}
    P_j = \dot{M}\eta_\phi\eta_a c^2\,,
\end{equation}
where the jet launching efficiencies are defined as:
\begin{equation}
    \eta_\phi = \left(\frac{\phi}{50}\right)^2\,;
    \eta_a = \left(1.063a^4+0.395a^2\right)\,,
\end{equation}
where $ \eta_a $ is the maximum efficiency for a given BH spin calibrated by \citet{Lowell2023}. In a MAD state $ \eta_\phi = 1 $, and thus Eq.~\eqref{eq:Pj} shows that the jet launching efficiency depends only on $ a $. This implies that the lGRB timescale is governed either by the BH spin-down timescale, $ \dot {a} $ \citep{Jacquemin-Ide2023}, or by the accretion timescale \citep[e.g.,][]{Gottlieb2022a}.

In contrast to collapsars, where the newly-formed BH is embedded in a dense massive stellar core, binary mergers take place in a considerably less dense environment surrounding the central engine. Consequently, jets can emerge well before the disk reaches a MAD state at $ \tmad $. Numerical simulations incorporating self-consistent models of binary mergers, capable of launching these jets, have verified this expectation \citep[e.g.,][]{Hayashi2022,Hayashi2023}. These simulations show that the compactness of the post-merger disk allows for the dimensional magnetic flux to rapidly accumulate on the BH\footnote{If the disrupted NS has a purely toroidal field configuration, $ \Phi $ is not constant, but slowly increases due to the dynamo process.}, resulting in a constant jet power, $ P_j(t<\tmad) \sim \Phi \sim {\rm const} $ (Eq.~\eqref{eq:BZ}). Due to the decaying mass accretion rate, the dynamical importance of the magnetic field (as measured by the dimensionless magnetic flux $\phi \propto \Phi \dot{M}^{-1/2}$) grows with time. Once $\phi \approx 50$ is reached, the disk enters a MAD state, which saturates the jet launching efficiency $\eta_\phi \approx 1$.  Thereafter, the jet power follows the declining mass accretion rate, $P_j (t>\tmad) \propto \dot{M}$ following Eq.~\eqref{eq:Pj}. 


Unlike collapsars, the disks formed from binary mergers do not have an external supply, resulting in their steady depletion and a continuous decrease in the BH mass accretion rate. In fact, at $ t \gtrsim 0.1\,\s $, the mass accretion rate $\dot{M}$ follows a single power-law decay without a characteristic timescale relevant to \cbGRBs (which in the collapsar case is set by the structure of the progenitor star). This implies that, in contrast to lGRBs where jet launching persists during the MAD phase of the disk and its timescale is set by $ \dot{M} $ or $ \dot{a} $, in mergers\footnote{The post-merger disk mass is negligible compared to the BH mass, so no appreciable spin-down is expected in binary mergers.} it is the MAD transition at $ \tmad $ (dictated by $ \mdisk $ and $ \Phi $) that eventually causes the jet power to decay, thus setting the \cbGRB duration, as we now describe.


\section{LBGRBs from BHs with massive disks}\label{sec:massive}

\citet{Gottlieb2023b} presented first-principles simulations of a BH--NS merger with mass ratio $ q = 2 $, which results in a rapidly spinning BH with $ a \simeq 0.86$.  A substantial accretion disk of mass $ \mdisk \approx 0.15\,\msun $ formed around the BH, resulting in a high initial accretion rate $ \dot{M} \sim M_\odot\,\s^{-1}$. We find a similar outcome here for five simulations of a BNS merger of component masses $ 1.06\,\msun $ and $ 1.78\,\msun $, initialized from the endpoint of the merger simulations of~\citet{Foucart:2022kon}. In that system, the remnant promptly collapses to a BH with $ a = 0.68 $, surrounded by a disk with $ \mdisk \approx 0.1\,\msun $ (see Appendix \S\ref{sec:simulations} for the full numerical results of the BNS merger simulations). 
Additionally, we perform five BH-NS merger simulations with component masses $ 4.05\,\msun $ and $ 1.35\,\msun $, respectively. The BH has a pre-merger spin of $a \approx 0.087$. The post-merger BH has a spin of $ a \approx 0.59 $ and mass of $ 5.26\msun $ and is surrounded by a disk with $ \mdisk \approx 0.007\,\msun $. Refer to Appendix \S\ref{sec:simulations} for a brief discussion of the simulation setup and time evolution of $ \dot{M},\phi,P_j,\eta_a \eta_\phi$. A detailed analysis of the results will be published in future work.

\begin{figure*}
    \centering
    	\includegraphics[width=7in]{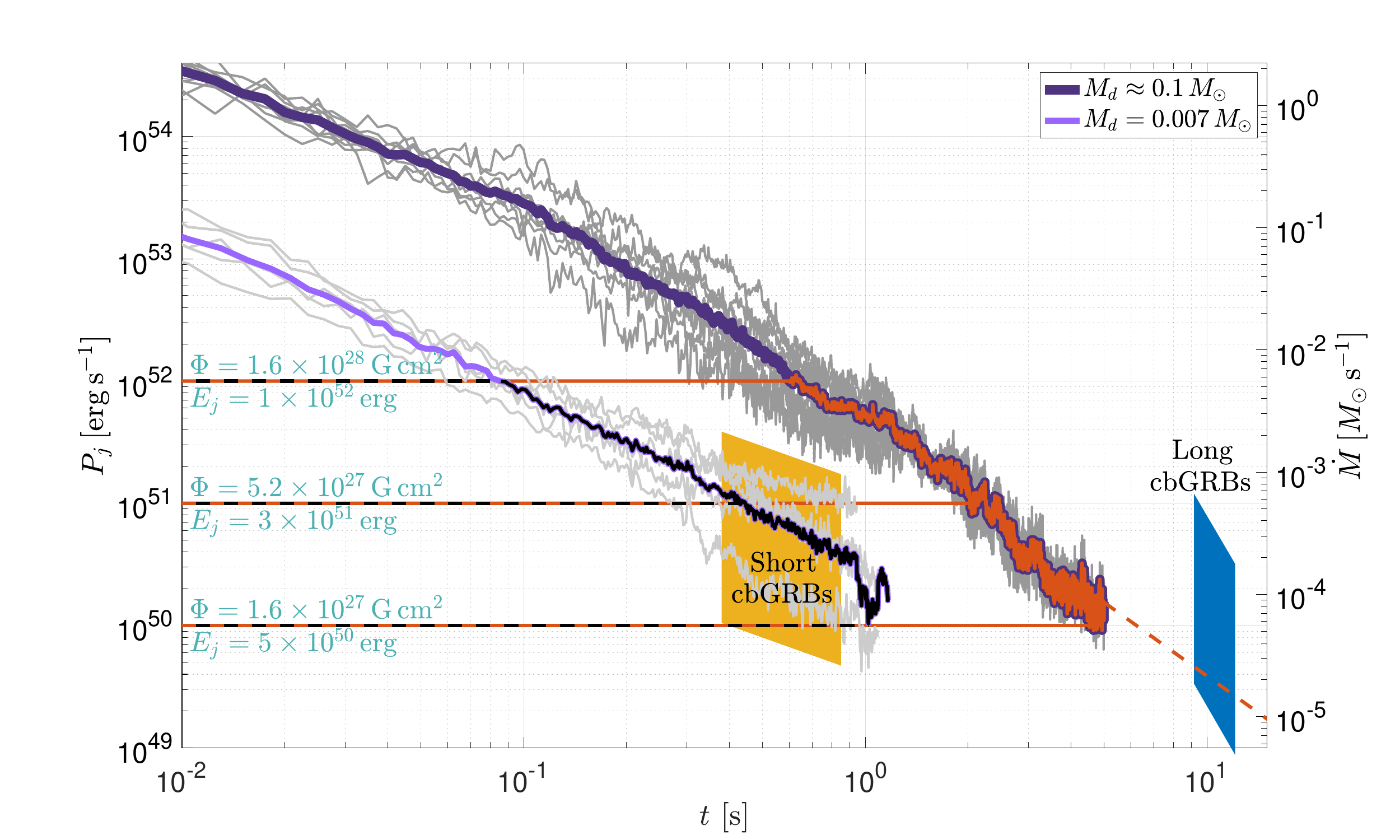}
     \caption{
     The jet power evolution of post-merger accretion disks for varying levels of magnetic flux ranging from non-MAD to MAD.
     Dark gray lines show the post-merger mass accretion rate evolution (right vertical axis) obtained for 4 BH--NS merger simulations \citep{Gottlieb2023b} and the 5 BNS merger simulations presented here, all of which generate massive disks $ \mdisk \approx 0.1\,\msun $. Light gray lines delineate the post-merger accretion rates from 5 BH-NS merger simulations that result in disk masses $ \mdisk \approx 0.007\,\msun $. The dark (light) purple lines mark the logarithmic averages of the mass accretion rates for $ \mdisk \approx 0.1\,\msun $ ($ \mdisk = 0.007\,\msun $), which constitute the maximum jet power assuming $\eta_a = 1$ corresponding to a BH spin $a \approx 0.87$ (left vertical axis). Black (for low mass disks) and orange (for high mass disks) lines illustrate schematically the jet power evolution for different assumptions about the dimensional magnetic flux threading the BH, $ \Phi $, and the corresponding total jet energy, $ E_j $ for the case of the massive disk. Since the magnetic flux on the BH is likely accumulated early and hence remains nearly constant before the disk transitions to MAD, the jet power, $ P_j $, is also predicted to be roughly constant at these times, powering the prompt emission. Once the dimensionless magnetic flux saturates in the MAD state, the jet power saturates at $P_j = \dot{M}c^2$ and thus follows the mass-accretion rate $\dot{M} \propto t^{-2}$ thereafter, powering the EE (we have extrapolated $P_j$ by a dashed line to later times). The yellow (blue) region outlines the estimated average jet power and duration $ T_{90} $ ($ T_{50} $) of the \sbGRB (\lbGRB) population based on prompt emission and afterglow observations (see text). While the jets from massive disks (orange lines) are either too powerful or operate for too long, compared to the prompt \sbGRB population, BH accretion from massive disks nicely matches the observed properties of prompt \lbGRBs. Jets from less massive disks (black lines) fit the luminosity and duration of \sbGRBs and are unable to give rise to \lbGRBs (see Figs.~\ref{fig:BNSsim},\ref{fig:BHNSsim} for the jet power evolution in simulations).
     }
     \label{fig:energetics}
    \end{figure*}

Eq.~\eqref{eq:Pj} shows that the jet power depends on both the mass accretion rate and the magnetic flux on the BH, $\Phi$. Binary compact mergers produce small accretion disks that promptly feed the available magnetic flux onto the BH\footnote{More massive BHs generally only lead to more compact disks \citep[e.g.,][]{Fernandez2020}, making this result robust.}. Because $ \Phi $ hardly changes thereafter during the subsequent accretion phase, this results in a constant jet power $ P_j \sim {\rm const} $ with a magnitude that depends on the disk's poloidal field strength. This is demonstrated in Figure~\ref{fig:energetics}, which depicts the jet power as a function of time for different values of $ \Phi $ and $ \mdisk $.

If the initial plasma beta in the disk is low (leading to large $\Phi$), then the jet launching efficiency is high, and the jet starts with too much power compared to prompt \sbGRB luminosities. In such cases, the dimensionless magnetic flux on the BH quickly saturates and the disk becomes MAD, ending the constant jet power phase. This translates to a relatively short and exceedingly luminous prompt \cbGRB (see e.g. the top black-red line in Fig.~\ref{fig:energetics}). This outcome challenges the model of \citet{Gao2022}, which suggests that a strong magnetic field can halt accretion to prolong the \cbGRB duration.

If instead, the initial plasma beta in the disk is high (low $\Phi$) or the initial magnetic field configuration is predominantly toroidal (see e.g., Appendix \S\ref{sec:simulations}), then the jet launching efficiency is low and the jet can generate a luminosity characteristic of prompt \sbGRBs. Over time, the efficiency increases due to the development of a global poloidal magnetic field and the decrease in the mass accretion rate that follows\footnote{Energy injection from alpha-particle recombination can also act to steepen the mass accretion power-law, after neutrino cooling is no longer important, at $ t \gtrsim 1\,\s $ \citep{Metzger2008,Haddadi2023}.} $ \dot{M} \sim t^{-2} $, as was also found in other numerical simulations \citep{Fernandez2015,Fernandez2017,Fernandez2019,Christie2019,Metzger2021,Hayashi2022}, where the normalization of the mass accretion rate is set by $ \mdisk $. When the disk finally becomes MAD at $ \tmad $, the efficiency stabilizes at $ \eta_\phi \approx {\rm const} $, and Eq.~\eqref{eq:Pj} reads $ P_j \sim \dot{M} \sim t^{-2} $. The two phases of $ P_j(t<\tmad) \sim P_0 $ and $ P_j(t>\tmad) \sim t^{-2} $ are generic for BH-powered \cbGRB jets. This motivates future analytic and numerical models to consider such temporal evolution of the jet power, with two free parameters: $ \tmad $, determined by the values of $ \phi $, and $ P_0 $, determined by $ \Phi $.

We stress that a roughly constant jet power does not imply a constant $ \gamma $-ray luminosity. Firstly, as shown in Fig.~\ref{fig:BNSsim}(d) in Appendix \ref{sec:simulations}, the jet power itself exhibits temporal variability, particularly for the initially toroidal configurations, owing to the stochastic nature of the dynamo process. Secondly, different portions of the jet undergo different levels of mixing and mass entrainment by the surrounding environment, leading to fluctuations in the baryon loading, magnetization, and Lorentz factor. These variations likely translate to a range of radiative efficiencies. This implies that even though the jet power remains roughly constant on average (consistent with the observed lack of temporal evolution in the statistical properties of GRB light curves throughout the burst; e.g., \citealt{McBreen2002}), different light curves can exhibit different shapes and variability, depending on the specifics of the merger.

\subsection{Constraints from cbGRB observations}

To compare the predictions of numerical simulations with observational data, we need to deduce the true jet properties from observations. The observed duration of the $ \gamma $-ray prompt emission from \cbGRB, $T_{90}$, varies depending on the detectors used \citep{Bromberg2013}, and whether the GRB duration distribution is modeled assuming 2 (lGRB and \cbGRB) or 3 (lGRBs, \sbGRBs, and \lbGRBs) populations. To estimate the range of $ T_{90}$ for \sbGRBs, we refer to the lowest and highest $ T_{90}$ values found among 2 and 3 Gaussian fits to Fermi and BATSE duration distributions in \citet{Tarnopolski2016} and find: $ 0.38\,\s \leq T_{90} \leq 0.85\,\s $. For \lbGRBs, we take the prompt emission durations of recent events GRB 211211A and GRB 230307A as boundaries: where $ T_{50} = 12.1\,\s $ \citep{Tamura2021} and $ T_{50} = 9.2\,\s $ \citep{Svinkin2023}, respectively. The use of $T_{50}$ instead of $T_{90}$, in this case, is motivated by the comparable radiated energies of the prompt burst and EE phases \citep{Kaneko2015,Zhu2022}, rendering $T_{50}$ a more accurate estimate for the prompt duration.

The characteristic jet power of \cbGRBs can be estimated as:
\begin{equation}
    P_{\rm obs} = \frac{f_b\Eisog}{\epsilon_\gamma T_{90 (50)}},
\end{equation}
where $\Eisog$ is the isotropic equivalent $ \gamma $-ray energy, $f_b$ is the beaming fraction, and $\epsilon_\gamma$ is the radiative efficiency, of the $ \gamma $-ray emission. We take $ \Eisog \approx 2\times 10^{51}\,\erg $ for \sbGRB \citep{Fong2015}, while for \lbGRB we adopt values $\Eisog \approx 5.3\times 10^{51}\,\erg$ \citep{Yang2022} and $\Eisog \approx 1.5\times 10^{52}\,\erg$ \citep{Levan2023b} measured for GRB 211211A and GRB 230307A, respectively.  We adopt a range of beaming factors $ 0.01 \leq f_b \leq 0.11 $ \citep{Fong2015}, corresponding to a true $ \gamma $-ray jet energy for \sbGRB of $ E_{\rm obs,\gamma} \approx 2\times 10^{49}-2\times 10^{50}\,\erg $ \citep{Fong2015}. Early estimates of the $ \gamma $-ray efficiency in lGRBs found $ \epsilon_\gamma \approx 0.5 $ \citep{Panaitescu2002}, but later analyses by \citet{Beniamini2015,Beniamini2016} suggested a lower value of $ \epsilon_\gamma \approx 0.15 $. \citet {Berger2014} found that the ratio of \cbGRB prompt to afterglow energy is higher by an order-of-magnitude compared to lGRBs, indicating a potentially higher $ \epsilon_\gamma $ for \cbGRBs. Nevertheless, this discrepancy might be attributed to the brighter afterglow emission arising from the denser large-scale environments surrounding the massive star progenitors of lGRBs.  It thus remains unclear whether the difference between lGRBs and \cbGRBs results from variations in the external medium, or is intrinsic (i.e., attributed to higher $ \epsilon_\gamma $ in \cbGRBs) due to e.g. substantial wobbling jet motion in collapsar jets \citep{Gottlieb2022d}. We thus consider a range of $ 0.15 \leq \epsilon_\gamma \leq 0.5 $ in our estimates.

Figure~\ref{fig:energetics} compares theoretical and numerical estimates of the jet power with \cbGRB observations. The right vertical axis shows the characteristic evolution of the BH accretion rate as a function of time after the merger (purple lines), which we have obtained by averaging the results of BH--NS merger and BNS merger simulations (gray lines), which produce massive disks with $ \mdisk \approx 0.1\,\msun $ (dark purple) and $ \mdisk = 0.007\,\msun $ (light purple). The jet power, displayed on the left vertical axis, is expected to be roughly constant at early times, insofar that most of the magnetic flux $ \Phi $ accumulates on the BH quickly. However, as the accretion rate drops, the dimensionless magnetic flux $\phi \propto \dot{M}^{-1/2}$ increases with time, until the disk enters a MAD state and the jet efficiency $ \eta_\phi \approx 1$ saturates, marking the characteristic MAD timescale, which represents the end of the prompt emission phase. After this point, the jet power $P_j \approx \eta_a\dot{M}c^2$ (Eq.~\eqref{eq:Pj}) tracks the decaying mass-accretion rate $P_j \propto t^{-2} $, which, as we show in \S\ref{sec:ee}, represents the EE.

As mentioned in \S\ref{sec:massive}, if the initial $ \Phi $ is high (top black-red lines), the jet is too powerful to match the characteristic power of \sbGRBs (yellow region) and \lbGRB (blue region). In order to achieve that power, the magnetic flux needs to be $ \Phi \sim 10^{27.5}\,{\rm G\,cm^2} $ (bottom orange line). For such a flux, if the disk is massive (dark purple), the accretion disk can only enter a MAD state after several seconds, significantly longer than the \sbGRB duration, $ \tmad \gg T_{90} $. On the other hand, flux at roughly this same level $ \Phi \lesssim 10^{27}\,{\rm G\,cm^2}$ leads to a jet which naturally achieves both the correct power and duration of the \lbGRB class (blue region, see Fig.~\ref{fig:BHNSsim} for the \lbGRB jet power evolution in simulations). Lighter disks (light purple) can enter the MAD state on the \sbGRB characteristic timescale (middle black line) to reproduce both the duration and luminosity of \sbGRBs (yellow region, see Fig.~\ref{fig:BHNSsim} for the \sbGRB jet power evolution in simulations).

We conclude that for relatively high disk masses $ \mdisk \gtrsim 0.1\,\msun $ (consistent with that required to produce the kilonova ejecta in GW170817; e.g., \citealt{Perego2017,Siegel2017}), the resultant jets exhibit either excessively high power (if the seed magnetic flux threading the disk is large) or lower power with extended duration of activity (if the seed flux is weaker). The former is ruled out observationally, implying that massive disks must give rise to \lbGRBs. Therefore, if the jet in GW170817 was powered by a BH surrounded by a massive disk, then the inferred jet energy, $ E_j \approx 10^{49}-10^{50}\,\erg $ \citep{Mooley2018b} indicates that the jet was not a luminous \cbGRB but rather a \lbGRB (e.g., the bottom orange line in Fig.~\ref{fig:energetics}). Unfortunately, because the jet was $ \sim 20^\circ $ off-axis \citep{Mooley2018b}, the bulk of the gamma-ray emission was beamed away from Earth, precluding a direct measurement of the jet duration.

\subsection{Disfavored solutions}\label{sec:solutions}

Here we explore potential caveats to the conclusions of the previous subsection. However, finding reasons to disfavor each, we shall ultimately conclude that BHs surrounded by massive disks remain the most likely explanation for \lbGRBs.

\subsubsection{Lower post-merger BH spins}

According to Eq.~\eqref{eq:Pj}, one potential way to reduce the jet power is to decrease the maximum efficiency $\eta_a$ by considering a lower post-merger BH spin for an otherwise similar magnetic flux. For example, a BH spin of $ a \approx 0.4 $ yields maximum efficiency of only $ \eta_a \approx 0.1 $ \citep{Lowell2023}. This would allow BHs with massive disks to power \sbGRBs provided the BH spin obeyed $a \lesssim 0.4$.  However, this requirement conflicts with the results of numerical relativity simulations, which find post-merger BH spins $0.6 \lesssim a \lesssim 0.8$ \citep{Kiuchi2009,Kastaun2015,Sekiguchi2016,Dietrich2017} for BNS mergers, corresponding to $0.3 \lesssim \eta_a \lesssim 0.7$. BH--NS mergers result in comparable or slightly higher remnant BH spins, at least for systems leading to the formation of massive accretion disks \citep{Foucart2011,Foucart2013,Foucart2014,Foucart2017,Foucart2019,Kyutoku2011,Kyutoku2015,Kawaguchi2015}. Appealing to a lower BH spin can thus only reduce the jet energy by a factor of $ \approx 2 $ compared to our estimates assuming $\eta_a \approx 1$.

\subsubsection{Delayed jet launching}\label{sec:delayed_launching}

As the magnetic field in post-merger accretion disks is anticipated to be predominantly toroidal \citep[e.g.,][]{Ruiz2018}, a jet of significant power may only be launched after a dynamo process in the disk generates a sufficiently strong global poloidal field.  If the seed magnetic field is weak, the jet onset might be delayed for several seconds \citep[see e.g.,][]{Hayashi2023}, thus operating for only a brief period before the disk transitions into a MAD state. This would make it possible for a BH with a massive disk to produce a \sbGRB. Nevertheless, it is unlikely that this scenario can serve as a generic explanation for \sbGRBs, as fine-tuning is required to launch the jet only briefly after $ \sim 10\,\s $, just before the disk reaches a MAD state, in order to achieve $ T_{90} \lesssim 1\,\s $.

\subsubsection{Misestimating the cbGRB duration}

Another possible caveat worth exploring is whether the jet duration could be inferred incorrectly from observations. Such an erroneous estimation could occur while (i) converting from the engine activity duration to $ T_{90} $, or (ii) due to uncertainties in observations:

(i) If the interaction of the jet with the external medium is sufficiently strong to decelerate the jet head to sub-relativistic velocities, the radial extent of the jet can become significantly shorter than $\tmad/c$, leading to an observed GRB duration considerably shorter than the MAD timescale over which the jet is launched. However, for typical properties of merger ejecta and \cbGRB jet energies, the jet head exhibits at least mildly relativistic motion from the onset \citep{Gottlieb2022f}, supporting the usual assumption that the GRB duration follows the activity time of the jet (i.e., $T_{90} \sim \tmad $).

(ii) In collapsars, the physics of jet propagation \citep{Bromberg2011b} and the observed GRB duration distribution \citep{Bromberg2012} support a substantial fraction of jets being choked inside the star \citep[see also][]{Gottlieb2022a}. Some jets may operate just long enough to break out of the star and power a short-duration GRB \citep{Ahumada2021,Rossi2022}. If collapsar jets outnumber those originating from binary mergers within the sGRB population, this could in principle lead to underestimates of the typical duration of binary merger jets.  However, while such an increase in the inferred $T_{90}$ of binary merger jets could potentially alleviate the tension in accounting for \sbGRB from massive BH disks, it provides no natural explanation for the bimodal distribution of GRB durations.

\section{Origin of prompt CBGRBs}\label{sec:origin}

While we conclude that massive disks likely produce \lbGRBs, Eq.~\eqref{eq:Pj} shows that \sbGRBs could instead emerge naturally from less massive BH disks. To explore whether variations in the disk mass from different merger outcomes are compatible with such a scenario, we now review the outcomes of compact binary mergers, as predicted by numerical relativity simulations of BNS
\citep{Rezzolla2010,Hotokezaka2011,Hotokezaka2013,Sekiguchi2011,Giacomazzo2013,Kiuchi2014,Kiuchi2015,Dietrich2015,Dietrich2017,Kastaun2015,Foucart2016,Kawamura2016,Sekiguchi2016,Hanauske2017,Radice2018b,Ruiz2018,Shibata2019} and BH--NS \citep{Shibata2006,Shibata2007,Shibata2008,Shibata2011,Etienne2008,Rantsiou2008,Duez2010,Foucart2012a,Foucart2011,Foucart2012b,Foucart2014,Foucart2017,Foucart2019,Kyutoku2011,Kyutoku2013,Kyutoku2015,Kawaguchi2015,Hayashi2021} mergers.

\subsection{Prompt-collapse Black Holes}\label{sec:bh}

When the total mass of a BNS exceeds a critical threshold $ \Mtot \gtrsim 2.8\,\msun $, the remnant created by the merger promptly collapses into a BH surrounded by an accretion disk \citep{Bauswein2013}, the mass of which depends sensitively on the binary mass ratio. For unequal mass ratios ($q \gtrsim 1.2 $), as characterized by our BNS merger simulations, the lighter NS is disrupted, resulting in a massive accretion disk, $ \mdisk \approx 0.1\,\msun $. By contrast, prompt-collapse mergers with $ q \approx 1 $ generate significantly smaller disk masses, $ \mdisk \lesssim 10^{-2}\,\msun $ \citep[see][for a review]{Shibata2019}. As the accretion rate scales linearly with the disk mass (Fig.~\ref{fig:energetics}), if $ \Phi $ is largely independent of $ \mdisk $, then disk masses of $ \mdisk \lesssim 10^{-2}\,\msun $ could power jets consistent with \sbGRB observations. This implies that \sbGRB can in principle be powered through massive BNS mergers with $ \Mtot \gtrsim 2.8\,\msun $ and $ q \approx 1 $. In BH--NS mergers, similarly low disk masses of $ \mdisk \lesssim 10^{-2}\,\msun $ are possible for high binary mass ratios, $ q \gg 1 $, low \emph{pre}-merger BH spin, or large spin-orbit misalignment~\citep{Foucart2018}.

    \begin{figure*}
    \centering
    	\includegraphics[width=7in,trim={0 0cm 0cm 0cm}]{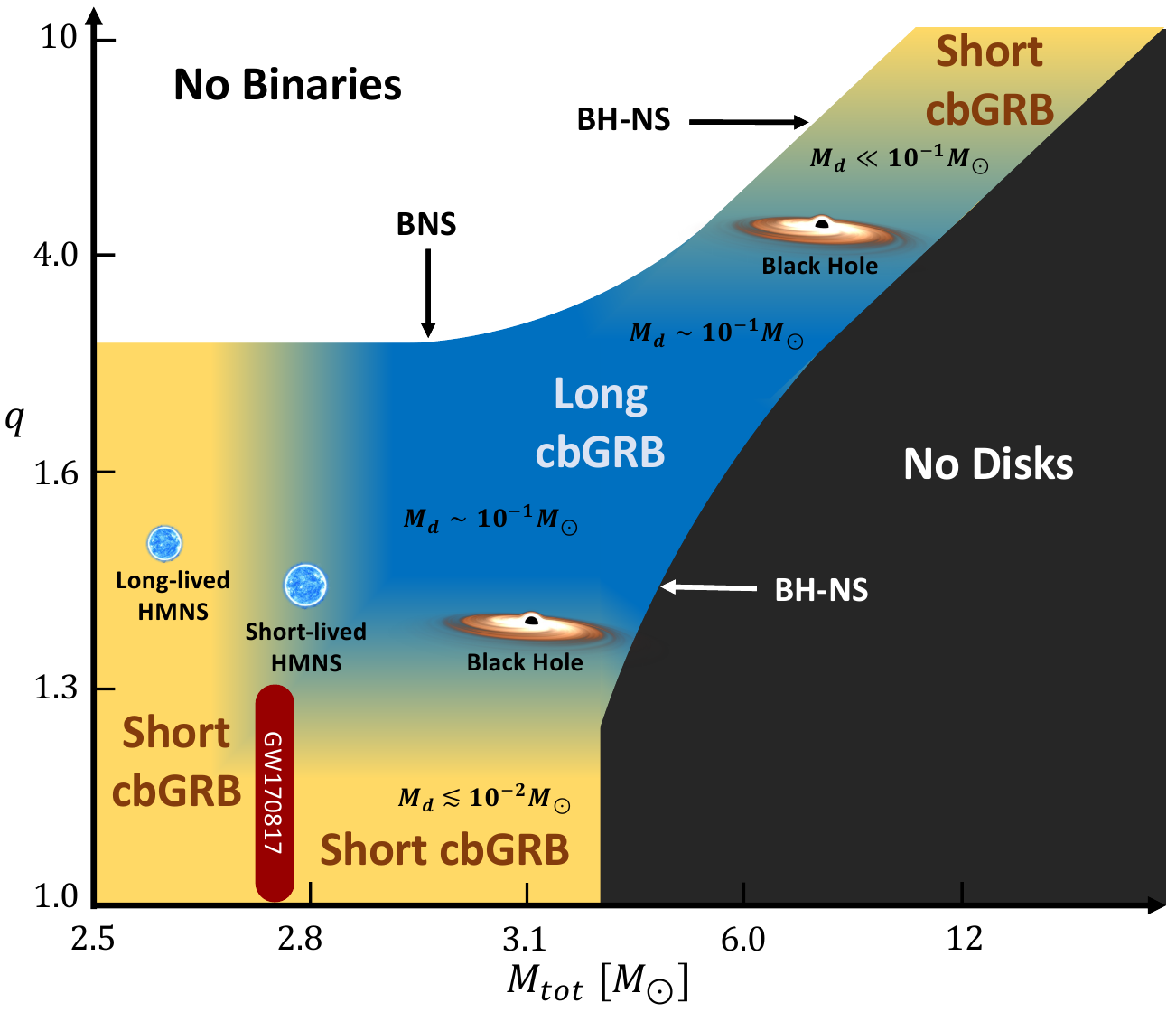}
     \caption{The outcomes of compact object mergers and their ability to power various \cbGRBs sub-classes as a function of the binary mass ratio (vertical axis) and total mass (horizontal axis). \lbGRBs occur in high $ \Mtot $ and high $ q $ BNS mergers that form a massive BH disk of $ \mdisk \sim 10^{-1}\,\msun $, or in high pre-merger BH spin and low mass ratio BH--NS mergers (blue region). \sbGRBs may arise either from equal mass ratio BNS mergers (bottom yellow region) and low pre-merger BH spin/high mass ratio BH--NS mergers (top yellow region), or by HMNS formed in BNS mergers with $ \Mtot \lesssim 2.8\,\msun $ (left yellow region). If BH-powered jets are different than HMNS-powered jets, then the absence of evidence for distinct sub-classes of \sbGRBs suggests that either BHs or HMNSs are likely to be the sole origin of these events, i.e. only one of the proposed \sbGRB scenarios is correct. The Galactic BNS mass distribution, the bimodal GRB duration distribution, and GW170817 observations favor HMNSs as the engine of \sbGRB jets.}
     \label{fig:sketch}
    \end{figure*}

The region $ \Mtot > 2.8\,\msun $ in Figure~\ref{fig:sketch} overviews this scenario. Low disk masses, such as those produced by equal mass BNS mergers that undergo prompt BH formation (bottom yellow region) or high mass ratio BH--NS mergers (top right yellow region)\footnote{Fig.~\ref{fig:sketch} should ideally cover the 3D space $(\Mtot,q,a)$, as the final disk mass is sensitive to the component of the initial BH spin aligned with the orbital angular momentum. Larger values of the BH spin result in more massive disks, and lower values of the BH spin in lower mass disks (or no disks at all). Nonetheless, the figure captures qualitatively the dependence of the results on $(\Mtot,q)$.}, giving rise to \sbGRBs. The opposite case of mergers forming massive BH disks then power \lbGRBs (blue region). If BHs power all \cbGRB jets, then it is expected that the \cbGRB duration spectrum will be continuous via the disk mass distribution. This seems to be in tension with the observed bimodal distribution. This scenario also poses an additional requirement on the rates given that most \cbGRBs arise from BNS mergers. If \sbGRBs are more common than \lbGRBs, this would require that $q \approx 1$ BNS mergers (\sbGRBs) should be more common than unequal mass ratio BNS mergers (\lbGRBs).  While consistent with the mass ratio distribution of the Galactic BNS population being narrowly concentrated around $ q \lesssim 1.2 $ \citep{Vigna2018,Farrow2019}, this picture is in tension with the BNS masses being below the expected prompt collapse threshold $\approx 2.8\,\msun$, as we now discuss.

\subsection{Long-lived HMNSs}\label{sec:hmns}

Observations of Galactic BNSs indicate an average NS mass of $ \mns \approx 1.33\,\msun$ \citep{Ozel2012,Kiziltan2013,Ozel2016,Farrow2019}. If representative of the extragalactic merger population as a whole, this relatively low mass suggests that most mergers will not undergo a prompt collapse into a BH given current constraints on the NS Equation of State (EoS) (e.g., \citealt{Margalit2019}). Furthermore, larger Fe cores are generally expected to result in both more energetic explosions and greater NS natal kicks, resulting in a correlation between these two properties \citep{Tauris2017}.  Since large kicks tend to unbind the binary, this makes less massive BNS systems more likely to eventually merge compared to their more massive counterparts.

The merger of BNS systems with $ \Mtot \lesssim 2.8\,\msun $ results in the formation of a highly magnetized differentially rotating HMNS, which only collapses into a BH after some delay \citep[e.g.,][]{Shibata2006b,Kastaun2015,Hanauske2017}. As a result of amplification of the magnetic field via differential rotational and instabilities, such HMNSs have the potential to produce energetic jets that could be the source of \sbGRBs \citep{Kluzniak1998}. One challenge to this scenario is that the polar outflows from HMNS are subject to baryon contamination of $ \sim 10^{-4}\,\msun\,{\rm str}^{-1} $ driven by strong neutrino heating from the atmosphere just above the surface \citep{Thompson2001,Dessart2009,Metzger2018}, which for jets of \sbGRB energies limits their bulk Lorentz factors to $ \Gamma \lesssim 10 $ \citep{Metzger2008b}. While relatively low, $ \Gamma \sim 10 $ might be nevertheless compatible with constraints based on compactness arguments in \cbGRBs \citep[][]{Nakar2007}\footnote{While compactness arguments in sGRB 090510 imply an ultra-relativistic Lorentz factor \citep{Ackermann2010}, it was proposed that this sGRB may be a misclassified collapsar event \citep[e.g.,][]{Panaitescu2011}}.

Comparing the observed properties of \cbGRBs with the energy output and lifetime of HMNSs is challenging due to the sensitivity of the latter to several theoretically uncertain properties of the post-merger system. The lifetime of the HMNS is governed by various physical processes, including neutrino cooling and angular momentum transport, the timescales for which in turn depend on factors such as the strength of the remnant's large-scale magnetic field, the saturation level of various magnetohydrodynamic instabilities giving rise to turbulent transport, and the initial distribution of angular momentum and temperature \citep{Margalit2022}. The complexity of incorporating all of these physical processes into long-term simulations, on top of uncertainties in the EoS, renders the lifetimes of HMNSs highly uncertain \citep{Hotokezaka2013,Dietrich2017}.  

More massive binaries in general produce HMNSs with shorter lifetimes \citep{Shibata2006b,Bauswein2013}.  For binaries with $ \Mtot \approx 2.7\,\msun $ the HMNS lifetime is primarily governed by angular momentum transport and the specific EoS \citep{Hanauske2017}. For less massive HMNSs, the collapse is dictated by either angular momentum transport with a timescale of $ \thmns \sim 0.1\,\s $, or if the HMNS is partially thermally supported \citep{Hotokezaka2013,Kaplan2014}, by neutrino cooling with a timescale of $ \thmns \sim 1\,\s $ \citep{Sekiguchi2011}. 
The binary mass ratio also plays a role, with greater asymmetry resulting in a longer HMNS lifetime due to increased angular momentum support \citep{Dietrich2017}.

Numerical simulations of $ q \lesssim 1.2 $ binaries with $ \Mtot \gtrsim 2.7\,\msun $, which birth long-lived ($ \thmns \approx T_{90} $) HMNS with strong magnetic fields $ B \gtrsim 10^{15}\,{\rm G} $, found the latter capable of generating \sbGRB-like emission \citep{Ruiz2016,Ruiz2020,Ciolfi2019,Ciolfi2020,Mosta2020,Combi2023,Kiuchi2023}. On the other hand, \citet{Most2023} found for a similar magnetic field and binary mass that the jet emission is lower by several orders of magnitude compared to other simulations. Furthermore, the HMNS lifetime varies greatly among those simulations, from $ \thmns \sim 10\,\ms $ to $ \thmns \gtrsim 1\,\s $, demonstrating the uncertainty in the HMNS lifetime, even when similar magnetic fields and $ \Mtot $ are considered \citep{Ruiz2016,Ruiz2020,Ciolfi2019,Ciolfi2020,AguileraMiret2023,Most2023,Kiuchi2023}. The specific properties of the binary and the EoS, thus play a crucial role in determining the characteristics of HMNSs.

Perhaps the tightest constraint on the properties of HMNSs comes through the interpretation of the first multi-messenger BNS system, GW170817, characterized by $ \Mtot \approx 2.75\,\msun $ and $ q \lesssim 1.3 $ \citep{Abbott2019b}. GW170817 provided valuable insights into the EoS of dense matter \citep{Radice2018a}, and supported the existence of a transient HMNS phase \citep{Margalit2017,Shibata2017,Rezzolla2018}.  The large quantity of slow-moving ejecta inferred from the kilonova, argues against a prompt collapse of the BH but is consistent with the expectation of disk outflows from a merger accompanied by a HMNS phase. The low inferred abundance of lanthanides in the ejecta (e.g., \citealt{Kasen2017}) supports strong neutrino irradiation of the disk by the HMNS \citep[e.g.,][]{Metzger2014,Kasen2015,Lippuner2017}. These findings thus point towards the requirement of a sufficiently stiff EoS, capable of supporting the formation of a HMNS from the GW170817 merger with $ \Mtot \approx 2.75\,\msun $. The HMNS could have persisted for the Alfv\'{e}n crossing timescale of $\sim 1\,\s $ \citep{Metzger2018}, sufficiently long to power a \sbGRB. Based on a suite of merger simulations targeted towards GW170817, \citet{Radice2018b} found that the remnant indeed most likely possessed enough angular momentum to prevent a collapse and to form a long-lived HMNS, even for $ \Mtot \approx 2.75\,\msun $.

The region $ \Mtot < 2.8\,\msun $ in Figure~\ref{fig:sketch} summarizes this alternative scenario, in which \sbGRBs arise from transient jets powered by moderately long-lived HMNSs formed from relatively low-mass binaries (left yellow region).
In this scenario, all prompt-collapse BHs give rise to \lbGRBs, where dimensional analysis suggests that $ \mdisk $ determines the jet power (\S\ref{sec:main}).

\subsection{Delayed-collapse Black Holes}\label{sec:delayed}

In BNS mergers where the combined mass is $ \Mtot \lesssim 2.8\,\msun $, the collapse of the HMNS into a BH may introduce a delayed launching of BZ-jets, which could potentially contribute to the \cbGRB populations. When the BH formation is preceded by a transient phase of a HMNS, the disk mass depends on $ \thmns $. If the HMNS collapses within a few ms, the system evolves in a similar way to prompt-collapse BHs. A longer-lasting HMNS with $ \thmns \gtrsim 10\,\ms $ allows for a greater opportunity for the post-collapse disk to grow through angular momentum transport to $ \mdisk \approx 0.1\,\msun $ \citep[e.g.,][]{Hotokezaka2013}. However, a longer-lived HMNS also provides an opportunity for the disk to lose mass prior to the BH formation. The disk continuously expands due to viscous angular momentum transport by the differentially rotating HMNS and viscous heating by magnetorotational instabilities (MRI) in the disk. Once neutrino cooling becomes subdominant to viscous heating, the disk expels winds, thereby reducing its mass \citep[see, e.g.,][]{Siegel2018,Fernandez2019}. In cases where vigorous viscous heating prompts rapid expansion, a substantial portion of the disk mass might be lost within $ \thmns $ \citep{Fujibayashi2018,Fujibayashi2020}.

The post-HMNS collapse disk mass remains elusive due to uncertainties pertaining to variables such as the magnetic field and effective viscosity in the disk, $ \thmns $, and other contributing factors. Given the significant impact of the disk mass on determining the \cbGRB type, the role of delayed-collapse BHs remains uncertain\footnote{Notably, the disk mass ejection timescale may bear observable implications, as early mass ejection from the disk shortens the freeze-out time for the electron fraction. Therefore, in scenarios with intense viscous heating, the electron fraction equilibrium is lower  \citep{Fujibayashi2020}, enabling us to estimate the disk mass at $ \thmns $ from kilonova observations.}. Two possibilities exist: (i) If the disk mass is appreciably reduced by viscous heating prior to BH formation, then the BZ-jet might be less luminous compared to the preceding HMNS-powered jet that generated the \sbGRB. In such instances, the jets launched by delayed-collapse BHs could serve as sources of EE once they transition into the MAD state.
(ii) If the viscous heating is insufficiently strong to remove the bulk of the disk mass on $ \thmns $ timescale, the BH forms with a massive disk. As outlined in \S\ref{sec:bh}, such disks are likely to give rise to \lbGRBs. If this configuration characterizes the standard picture of HMNSs, the \lbGRBs would supersede the observational imprint of HMNS-powered jets, indicating that all \cbGRBs are powered by BHs. Interestingly, this perspective forecasts that BNS mergers with $ \Mtot \lesssim 2.8\,\msun $ lead to \lbGRBs, implying that \lbGRBs are more common than \sbGRBs.

\subsection{Long-lived SMNSs}\label{sec:smns}

For particularly low-mass binaries $ \Mtot \lesssim 2.4\,\msun $, a very long-lived rigidly rotating SMNS with $ \mdisk \approx 0.1\,\msun $ can form \citep{Giacomazzo2013,Foucart2016}. Similar to the HMNS case, the early stages after the formation of a SMNS can in principle give rise to moderately relativistic outflows with $ \Gamma \sim 10 $ (e.g., \citealt{Metzger2008b}). However, SMNSs can live for $ t \gg 1\,\s $ before collapsing, and thus may generate a relativistic wind that reaches $ \Gamma \gtrsim 100 $ as the rate of neutrino-driven mass-ablation from the SMNS surface decays (e.g., \citealt{Thompson2004,Metzger2008b}). Relativistic magnetohydrodynamic (MHD) \citep{Bucciantini2012} and numerical relativity \citep{Ciolfi2017,Ciolfi2020,Ruiz2020} simulations have demonstrated that long-lived magnetars are potentially capable of powering \cbGRB jets. Such jets could be compatible with energy injection into \cbGRB afterglows \citep{Zhang2001}, and the late-time spin-down luminosity of the magnetar obeys $\sim t^{-2}$, also consistent with the observed decay evolution of the EE \citep{Metzger2008b,Bucciantini2012,Gompertz2013}.

The kilonovae which accompanied the two recent \lbGRBs, GRB 211211A and GRB 230307A, support relatively slow outflows ($ \vej \lesssim 0.1c $) containing high-opacity material consistent with significant lanthanide/actinide enrichment \citep{Rastinejad2022,Levan2023b,Barnes2023}. While both these properties are consistent with the disk outflows from a BH accretion disk (e.g., \citealt{Siegel2017,Fernandez2019}), the ejecta velocities are too low compared to those expected following substantial energy injection from the magnetar wind \citep{Bucciantini2012}.  Sustained neutrino irradiation of the disk outflows from the hot stable neutron star remnant, also precludes significant heavy $r$-process material \citep[e.g.,][]{Metzger2014,Kasen2015,Lippuner2017}.  

Additional arguments which disfavor SMNSs as the progenitors of the majority of the \cbGRBs include: (i) lack of evidence for a significant injection of rotational energy from the magnetar based on the late radio afterglow emission \citep{Metzger2014b, Horesh2016, Schroeder2020, Beniamini2021}; (ii) the BNS mass distribution favors HMNSs as the common remnant of a BNS merger, and recent results by \citet{Margalit2022} show that accretion can shorten the SMNS lifetime such that it is closer to $ \thmns $, reducing the parameter space capable of generating long-lived magnetars. In light of the viability of the massive BH disk scenario, the above arguments disfavor the model suggested by \citet{Metzger2008b}, \citet{Sun2023}, in which \lbGRBs with EE are powered by long-lived magnetars.

\subsection{Binary WD merger and AIC}\label{sec:WD}

The formation of a magnetized NS does not require a merger that involves a pre-existing NS. Instead, it may originate from the gravitational collapse of a WD in a binary system \citep{Taam1986}. The secondary star for AIC can either be a merging WD companion, or a non-degenerate donor \citep[e.g.,][]{Duncan1992,Usov1992,Yoon+07}. The resulting newly formed NS can be a magnetar if the magnetic field of the progenitor WD is very strong and is amplified by flux freezing during the collapse \citep[see e.g.,][]{Burrows2007} or after the collapse through magnetic winding or other dynamo action after the merger/collapse. Magnetars formed from AIC may potentially act as central engines for \cbGRBs \citep{Usov1992,Metzger2008b}.

Accreting WDs are generally considered to lose much of their angular momentum during their evolution (e.g., through classical nova eruptions), ultimately becoming slow rotators \citep{Berger2005}. In the case of binary WD mergers, the angular momentum budget is much higher initially; however, the most massive mergers capable of undergoing AIC ultimately produce an NS with a mild rotation period of $ \sim 10\,\ms $, due to angular momentum redistribution during the post-merger phase prior to collapse \citep{Schwab2021}. Such slowly rotating magnetars have a limited energy reservoir and would not be accompanied by an appreciable accretion disk.

AIC occurs when a massive oxygen–neon WD accretes matter from a companion star until it reaches the Chandrasekhar limit and collapses into an NS (e.g., \citealt{Nomoto&Kondo1991}; however, see \citealt{Jones+16}). During the collapse process, conservation of angular momentum may lead to the formation of a rapidly spinning NS surrounded by a disk \citep{Bailyn1990}. Additionally, the fast and differential rotation in the newly formed NS results in a substantial amplification of the magnetic field \citep{Dessart2007}, which may result in a millisecond magnetar. However, the AIC faces similar challenges as the SMNS scenario (\S\ref{sec:smns}). For example, neutrino irradiation from the long-lived magnetar will increase the electron fraction in the disk outflows \citep[e.g.,][]{Metzger2009,Darbha2010}, leading to inconsistencies with the lanthanide-rich ejecta inferred from the kilonova emission from GRB 211211A and GRB 230307A.

Another scenario involving WDs is an NS--WD merger \citep{Fryer+99,King+07}, which was proposed as origins of GRB 211211A \citep{Yang2022} and possibly GRB 230307A \citep{Sun2023}.  It is argued that the burst duration scales with the accretion timescale, which in turn scales inversely with the density of the companion star for an accretion-powered engine, favoring a WD. However, as we have shown in \S\ref{sec:massive}, the burst timescale depends on the disk mass and the magnetic flux threading the BH and does not necessarily require a low-density WD to prolong the accretion timescale. In fact, we find that after $ t \sim 100\,\ms $, the mass accretion rate follows a single power-law profile, indicating that there is no accretion timescale relevant to \cbGRBs. Additionally, proton-rich matter accreted from the disrupted WD is unlikely to reach high enough densities to produce neutron-rich outflows capable of generating any significant $r$-process material, much less the relatively heavy lanthanides (\citealt{Metzger2012}; see \citealt{Fernandez2019b} for simulations of the post-merger disk evolution and nucleosynthesis). The NS--WD merger scenario thus faces difficulties in explaining the observed kilonova emission \citep[see][and references therein]{Barnes2023}.

\subsection{Neutrino annihilation}\label{sec:neutrino}

The high accretion rates anticipated in post-merger disks give rise to strong neutrino emission. Efficient annihilation of neutrinos and anti-neutrinos can generate relativistic jets that may power \cbGRBs \citep[e.g.,][]{Woosley1993}. These jets are expected to operate as long as the accretion rate is $ \dot{M} \gtrsim 10^{-2}\,\msun $ \citep{Popham1999}. This requirement implies that massive disks are necessary \citep[e.g.,][]{Leng2014} to enable jet launching for $ T_{90} \lesssim 1\,\s $. If the initial magnetic field in the disk is predominantly toroidal, then BZ-jet may follow the neutrino-driven jet after $ t \gtrsim 1\,\s $ \citep[e.g.,][]{Christie2019,Gottlieb2023b}, and power the late EE \citep{Barkov2011}. This scenario cannot explain \lbGRBs, and as we now argue, is also disfavored as the origin of \sbGRBs.

The main limitation of neutrino-driven jets lies in their available energy \citep{Leng2014,Just2016}. In BNS mergers, where a significant amount of ejecta is expected along the polar axis, these low-energy jets would fail to break out and generate a \cbGRB \citep{Just2016}. Furthermore, the mass distribution of the Galactic BNS population suggests that most post-merger remnants are HMNSs. The large amount of mass in the HMNS atmosphere (\S\ref{sec:hmns}) would load neutrino-driven jets with baryons, hindering their ability to achieve relativistic velocities \citep{Dessart2009}. Consequently, such jets would be incapable of producing \cbGRBs.


\section{Origin of the precursor flare and extended emission, and comparison of BH-powered and HMNS-powered jets}\label{sec:comparison}

    \begin{figure*}
    \centering
    	\includegraphics[width=7in,trim={0cm 0cm 0cm 0cm}]{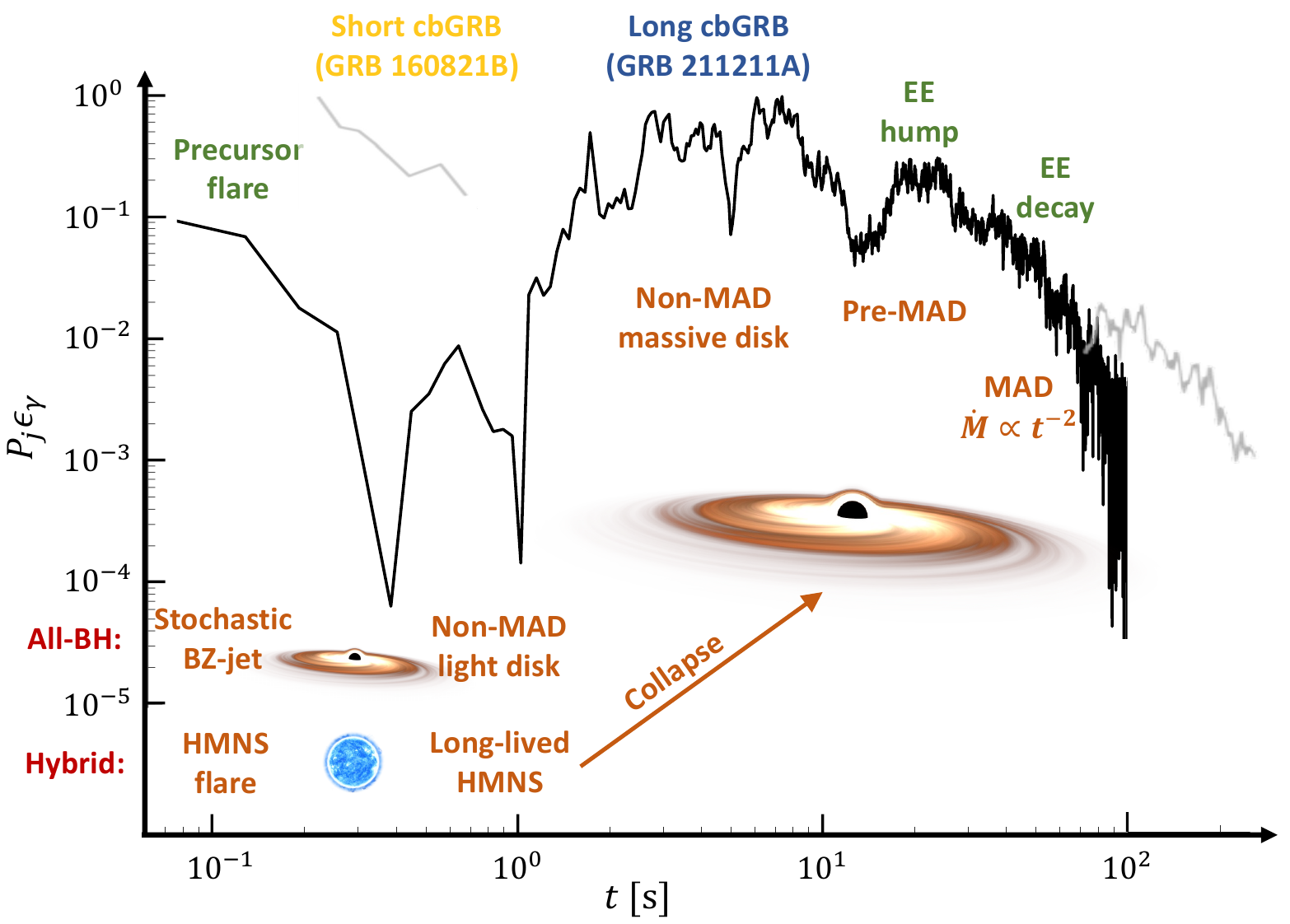}
     \caption{An illustration of how the underlying physics of the merger product (orange) in the hybrid and all-BH scenarios (red) translates into different phases in the \cbGRB light curves: \sbGRB (yellow), \lbGRBs (blue) and preceding and succeeding phases (green): 
     The precursor flare (\S\ref{sec:flare}) can be generated by either the accumulation of a stochastic magnetic field on a BH \citep{Gottlieb2023b} or by a HMNS \citep{Most2023}. A \sbGRB can be powered either by a BH surrounded by a non-massive disk before it transitions to a MAD state (\S\ref{sec:bh}), or by a long-lived HMNS (\S\ref{sec:hmns}). On the other hand, a \lbGRB emerges from BHs with massive disks before they enter the MAD state (\S\ref{sec:lgrb}), whether the BH formed promptly in the All-BH scenario, or followed a HMNS collapse in the Hybrid scenario. Finally, the BH disk becomes MAD and follows $ P_j \sim t^{-2} $ to power the EE.
     Representations of the light curves of the \lbGRB 211211A \citep{Rastinejad2022} and \sbGRB 160821B \citep{Stanbro2016} are shown in black and gray, respectively, in a log-log scale. Both are confirmed to be \cbGRBs by their detected kilonova counterparts \citep{Lamb2019,Rastinejad2022}
     }
     \label{fig:lightcurve}
    \end{figure*}

Figure~\ref{fig:lightcurve} utilizes the light curves of \lbGRB 211211A (black) and \sbGRB 160821B (gray) to illustrate the connection between the underlying physics of the compact object (orange labels) and the various phases observed in the \cbGRB light curve (yellow for \sbGRBs, blue for \lbGRBs, and green for preceding and succeeding phases). A \sbGRB can be powered either by a BH with a light accretion disk, or as inferred by the kilonova observations of GRB 160821B -- by a long-lived HMNS \citep{Lamb2019}, before collapsing into a BH. A \lbGRB is fueled by a BH surrounded by a massive accretion disk, as the dimensionless magnetic flux threading the BH steadily accumulates. The origin of the precursor flare and the EE are discussed below.

Up to this point, we have presented both HMNS-powered and BH-powered jets as potential contributors to \sbGRBs. However, there is no evidence indicating the existence of two distinct sub-populations among \sbGRBs, suggesting that only one of these engines is responsible for producing the majority of \sbGRBs. Table~\ref{tab:scenarios} summarizes the origin of \sbGRBs and \lbGRBs, as well as the outcomes of the different types of mergers, as predicted in both scenarios. We denote the scenario in which HMNSs power \sbGRBs and BHs power \lbGRBs by the ``hybrid'' scenario. The scenario in which all \cbGRBs are powered by BHs, with the GRB duration increasing with the disk mass, is denoted by ``all-BH'' scenario. Both scenarios predict the formation of a \lbGRBs when the BH is surrounded by a massive disk. When a less massive disk is present (in nearly-equal mass ratio BNS mergers with $ \Mtot \gtrsim 2.8\,\msun $, or in BH-NS mergers with either high $ q $ or low $ a $), the all-BH scenario predicts a \sbGRB signal, whereas the hybrid scenario predicts a \lbGRB signal. When $ \Mtot \lesssim 2.8\,\msun $, the \cbGRB duration in the all-BH scenario depends on the uncertain post-HMNS collapse disk mass (see \S\ref{sec:delayed}).

In the all-BH scenario, the \cbGRB duration spans a continuous spectrum, whereas, in the hybrid scenario, the BH-powered \lbGRBs comprise a separate class. Therefore, the hybrid scenario offers a natural distinction between \sbGRBs powered by HMNSs and \lbGRBs powered by BHs. Furthermore, the hybrid scenario finds support from the bimodal \cbGRB duration distribution, the mass distribution of BNS systems, as well as from observations and simulations of GW170817. In the following subsections, we show that the hybrid scenario is also more compatible than the all-BH scenario with all phases of the \cbGRB light curve.

	\begin{table}
		\setlength{\tabcolsep}{2pt}
		\centering
		\renewcommand{\arraystretch}{1.4}
		\begin{tabular}{| c c | c  c | }
  \hline
			  \multicolumn{2}{|c|}{\multirow{2}{*}{Event type}} & \multicolumn{2}{c|}{Scenario:} \\
			\multicolumn{2}{|c|}{} & Hybrid & All-BH
			\\	\hline
			\multicolumn{2}{|c|}{\sbGRB engine} & HMNS & BH + $ \mdisk \lesssim 10^{-2}\,\msun $\\
                \multicolumn{2}{|c|}{\lbGRB engine} & \multicolumn{2}{c|}{BH + $ \mdisk \approx 10^{-1}\,\msun $} \\\hline
                \multirow{4}{*}{BNS} & $ \Mtot \lesssim 2.8\,\msun; q\lesssim 1.2 $ & \sbGRB & \sbGRB \\
                & $ \Mtot \lesssim 2.8\,\msun; q\gtrsim 1.2 $ & \sbGRB & \lbGRB \\
                & $ \Mtot \gtrsim 2.8\,\msun; q\lesssim 1.2 $ & \lbGRB & \sbGRB \\
                & $ \Mtot \gtrsim 2.8\,\msun; q\gtrsim 1.2 $ & \lbGRB & \lbGRB
                \\\hline
                \multirow{3}{*}{BH-NS} & Low $ q $ \& high $ a $ & \lbGRB & \lbGRB \\
                & High $ q $ $ \oplus $ low $ a $ & \lbGRB & \sbGRB \\
                & High $ q $ \& low $ a $ & No GRB & No GRB
                \\\hline
                
		\end{tabular}
		
		\caption{
			Summary of the mapping between the Hybrid and All-BH scenarios and associated \cbGRB classes.
    		}
    		\label{tab:scenarios}
	\end{table}

\subsection{Precursor flare}\label{sec:flare}

Each of the proposed hybrid and all-BH scenarios postulates a different physical origin for the precursor flare before the rise of the main burst. In the hybrid scenario, \citet{Most2023} demonstrated how the differentially rotating HMNS builds loops with footpoints at different latitudes on its surface. The resultant twist in the loop causes it to become unstable, inflate and buoyantly rise, forming a bubble that is entirely detached from the HMNS surface, and erupting after reconnecting \citep[e.g.,][]{Carrasco2019,Mahlmann2023,Most2023}. This behavior powers quasi-periodic flares prior to the jet formation.

For BH-powered jets, \citet{Gottlieb2023b} showed that if the seed magnetic field in the disk is toroidal, as expected in binary systems, then the stochastic accumulation of incoherent magnetic loops on the horizon can lead to a short burst of energy (see model $ T_s $ in their figure 1(d)), which may constitute the precursor flare. As more flux reaches the BH, the stochastic field cancels out by virtue of contribution of loops of different polarity. Consequently, the total flux drops to zero, before starting to build a large-scale poloidal field through the dynamo process and power the \cbGRB emission. Due to the stochastic nature of the accumulated flux, the flare energy is expected to be very weak, and the resultant outflow may not be able to punch through the optically thick disk wind and/or dynamical ejecta \citep{Gottlieb2023b}. Therefore, the emergence of such precursor flares in the all-BH scenario may require fine-tuning. Nevertheless, it is possible that the precursor in the all-BH scenario is also powered by a short-lived HMNS before it collapses into a BH on a $ \sim 10\,\ms $ timescale.

\subsection{Main cbGRB burst}\label{sec:main}

Dimensional analysis suggests that $ \Phi \sim \sqrt{\mdisk} $, thus $ \mdisk \sim \Phi^2 \sim P_j^2 $, while the dimensionless magnetic flux $ \phi $ is independent of $ M_d $. This is also supported by the fact that the saturation level of the amplified ordered field in the disk seems to scale with the turbulent disk pressure, which in turn likely scales with $ \dot{M} $. This implies that reducing the disk mass results in a lower jet power, rather than shortening the \cbGRB duration, which scales with the \emph{dimensionless} magnetic flux (see \S\ref{sec:massive}). Namely, massive disks produce GRB 211211A-like \lbGRBs, whereas lower mass disks produce less luminous \lbGRBs, which are harder to detect. Therefore, unless there is an intrinsic correlation between $ \mdisk $ and $ \phi $, the variation in $ \mdisk $ does not naturally yield the variation in the \cbGRB duration. This favors BHs with less massive disks to power weaker \lbGRBs, and \sbGRBs as a distinct \cbGRB population, which emerges from HMNSs.


\subsection{Extended emission}\label{sec:ee}

Following the main hard burst, the softer EE phase commences. In both hybrid and all-BH scenarios, an accretion disk forms and is present at the time of the EE. Once the disk enters the MAD state, the jet power evolves in accordance with the mass accretion rate, $ P_j \sim t^{-2} $, similar to the observed temporal evolution of the EE decay. The preceding flat EE hump is thus generated by the constant power jet, just before the disk transitions to a MAD state. The EE may end once the disk is overheated after $ \sim 100\,\s $, and evaporates on this timescale \citep{Lu2023}. This evolution of a constant jet power followed by a $ t^{-2} $ decay for another order of magnitude in time naturally results in a comparable energy content between the \cbGRB prompt emission and the EE, as suggested by observations \citep{Kaneko2015}.


Any \cbGRB model must account for the reason why the EE likely emerges $ \sim 10\,\s $ after the onset of the prompt emission. This implies that if the EE follows a \sbGRB where $ T_{90} \ll 10\,\s $, there must be a quiescent period between the prompt and the EE phases \citep[e.g.,][]{Perley2009}. The all-BH scenario, which posits that both \cbGRB types are powered by BHs, encounters difficulties in explaining this constraint. As described in \S\ref{sec:massive}, BHs launch jets with a constant power followed immediately by the EE decay once the disk transitions to a MAD state. Therefore, no quiescent times would be expected to emerge between the prompt emission and the EE phase. In the hybrid model, \sbGRBs are powered by HMNSs and the post-collapse BZ-jet generates the EE. The time between the HMNS collapse and the launch of the BZ-jet offers a natural explanation for the occurrence of the observed quiescent interval.

\section{Conclusions}\label{sec:summary}

The discoveries of $ \sim 10 $-s long prompt emission in \lbGRBs 211211A \citep{Rastinejad2022} and 230307A \citep{Levan2023b}, followed by softer EE signals, suggest that the \cbGRB population can be divided into two classes: \sbGRBs ($ T_{90} \lesssim 1\,\s $) and \lbGRBs ($ T_{50} \sim 10\,\s $). However, the underlying physics that differentiates these classes and the origin of the prolonged EE are poorly understood. Moreover, drawing inferences about the astrophysical properties of binary mergers from \cbGRB observables poses a formidable challenge. In this paper, we have developed a novel theoretical framework that connects different binary merger types to the distinct sub-populations of \cbGRB and to the different components in their light curves. This provides the very first solution for the origin of both the constant-power prompt emission and decaying EE from first principles.

In collapsars, the presence of a dense stellar core surrounding the BH hinders the launching of jets when the accretion disk is not in a MAD state. This implies that for lGRBs, the jet operates in a MAD state at all times, and the characteristic lGRB duration can be set by either the mass accretion rate or by the BH spin-down timescale. By contrast, in binary systems where the environment is less dense, the conditions allow for the launching of the jet before the disk enters the MAD state. Due to the compactness of the disk, the dimensional magnetic flux, $ \Phi $, quickly accumulates on the BH, resulting in a roughly constant jet power before the transition to MAD occurs. After the accretion disk enters the MAD state, the jet power follows the mass accretion rate of $ P_j \sim \dot{M} \sim t^{-2} $, signaling the end of the prompt emission phase and the onset of the decaying EE. This behavior is consistently observed in all first-principles simulations and should be considered when modeling \cbGRB jets. In this jet power evolution model, there are two free parameters: (i) the time of the transition to a MAD state, which determines the \cbGRB duration and is influenced by $ \phi $; (ii) the magnitude of the constant jet power, which is governed by $ \Phi $.

The nature of the resultant central engine is determined by the total mass of the binary system. Unequal mass ratio BNS mergers with $ \Mtot \gtrsim 2.8\,\msun $ and BH--NS mergers with a moderate mass ratio and high pre-merger BH spin lead to the formation of a BH surrounded by a massive ($ \mdisk \gtrsim 0.1\,\msun $) accretion disk. Depending on $ \Phi $ (as illustrated in Fig.~\ref{fig:energetics}), such a massive disk can give rise to either extremely bright \sbGRB, or \lbGRB. Analyzing the \sbGRB and \lbGRB observational data, we conclude that massive disks inevitably power long-duration signals, and thus are most likely the progenitors of \lbGRBs such as GRB 211211A and GRB 230307A. Lighter disks with $ \mdisk \lesssim 10^{-2}\,\msun $ can produce typical \sbGRBs.

In other merger configurations, the resultant BH disk is less massive, and if $ \Phi $ is weakly dependent on $ \mdisk $, a \sbGRB jet can be generated. While this interpretation of \cbGRBs powered by BHs provides an explanation for \sbGRBs and \lbGRBs, it faces challenges in explaining various observational features in \cbGRB light curves, including flares observed before the prompt emission and the quiescent time observed between the prompt emission and the EE. Most importantly, the Galactic BNS population suggests that most binary systems have $ \Mtot \lesssim 2.75\,\msun $ \citep[e.g.,][]{Ozel2012,Kiziltan2013}, where a prompt collapse into a BH is not anticipated.

In BNS mergers with $ \Mtot \lesssim 2.8\,\msun $, the product of the merger is a HMNS \citep[e.g.,][]{Margalit2019}. Both analytic and numerical studies demonstrated that HMNSs are capable of generating relativistic jets that power \cbGRBs \citep[e.g.,][]{Metzger2008b,Kiuchi2023}. The best-studied event in this mass range is the multi-messenger GW170817 with $ \Mtot \approx 2.75\,\msun $. The associated kilonova signal observed in GW170817 supports the formation of a long-lived ($ \thmns \lesssim 1\,\s $) HMNS \citep[][]{Metzger2018,Radice2018a}. This timescale is sufficiently long to power \sbGRBs. Unlike BHs, HMNSs can naturally produce precursor flares \citep{Most2023}, and account for the quiescent time between the prompt and the EE by virtue of the transition from HMNS-powered to BH-powered jets. 

Various constraints, from kilonova observations to radio constraints on late-time rotational energy injection, favor prompt-collapse BH-powered jets and HMNS-powered jets over models that include long-lived magnetars, WDs, or neutrino-driven jets. While we thus find it likely that BHs with massive disks are responsible for \lbGRBs, we are less certain about the origin of the shorter \sbGRB population.  {\it A priori}, both BH-powered jets (BH--NS mergers or BNS mergers with $ \Mtot \gtrsim 2.8\,\msun $ and $ q \lesssim 1.2 $) and HMNS-powered jets ($ \Mtot \lesssim 2.8\,\msun $) remain viable possibilities (Fig.~\ref{fig:sketch} and Tab.~\ref{tab:scenarios}). However, the lack of evidence for two distinct sub-classes among the \sbGRB population, suggests that if HMNS-powered jets are different than BH-powered jets, then one of these channels dominates. We find several reasons to prefer transient HMNSs over low-disk mass BHs in this case.

A key distinction between the all-BH and hybrid scenarios lies in the \cbGRB duration distribution. BH-powered jets should exhibit a continuous spectrum from \sbGRBs to \lbGRBs, scaling with the binary mass ratio. Conversely, if HMNSs are the progenitors of \sbGRBs, they differ intrinsically from BH-powered \lbGRBs, proposing two distinct \cbGRB classes. The recent joint detections of \cbGRBs with kilonovae provide an exciting opportunity to assemble a sizable sample of confirmed \cbGRB events. Analyzing this collection could shed light on whether kilonova-associated \sbGRBs and \lbGRBs form a continuous spectrum or represent distinct classes. This, in turn, may enable us to deduce whether HMNSs, BHs, or both, serve as the primary progenitors of \sbGRBs.

\acknowledgements

We thank the referee, Alexander Tchekhovskoy, Rosalba Perna, Jonatan Jacquemin-Ide, Om Sharan Salafia, and Daniel Kasen for valuable discussions. We thank Jillian Rastinejad for providing the observational data for GRB 211211A.
OG is supported by Flatiron Research and CIERA Fellowships. OG acknowledges support by Fermi Cycle 14 Guest Investigator program 80NSSC22K0031, and NSF grant AST-2107839.
BDM acknowledges support from the National Science Foundation (grant number AST-2002577). DI is supported by Future Investigators in NASA Earth and Space Science and Technology (FINESST) award No. 80NSSC21K1851. An award of computer time was provided by the ASCR Leadership Computing Challenge (ALCC), Innovative and Novel Computational Impact on Theory and Experiment (INCITE), and OLCF Director's Discretionary Allocation programs under award PHY129. This research used resources of the National Energy Research Scientific Computing Center, a DOE Office of Science User Facility supported by the Office of Science of the U.S. Department of Energy under Contract No. DE-AC02-05CH11231 using NERSC awards ALCC-ERCAP0022634 and NP-ERCAP0020543 (allocation m2401). This research was facilitated by the Multimessenger Plasma Physics Center (MPPC), NSF grant PHY-2206610.

\bibliography{refs}

\appendix

\section{Binary merger simulations}\label{sec:simulations}

Our simulation setup is similar to that described in \citet{Gottlieb2023b,Gottlieb2023c}. Below we summarize the setup and main properties of the simulations.
We employ a numerical relativity simulation performed with the SpEC code \citep{spec} that evolves the system from the pre-merger phase to 10 ms after the prompt collapse to a BH.
In the BNS merger simulations, the masses of the merging NSs are $ 1.06\,\msun $ and $ 1.78\,\msun $, and the neutron stars are described with the LS220 equation of state~\citep{Lattimer:1991nc}. The merger product is a BH with mass $ \mbh = 2.67\,\msun $ and dimensionless spin $ a = 0.68 $. The BH is surrounded by a massive accretion disk of $ \mdisk = 0.096\,\msun $. Details of the SpEC simulation can be found in~\citet{Foucart:2022kon}. The merger simulation includes general relativity (GR), relativistic fluid dynamics, Monte-Carlo neutrino transport, and a subgrid viscosity model to approximate angular momentum transport and heating due to MHD instabilities.
In the BH-NS merger simulation, the mass of the NS is $ 1.35\,\msun $ and the mass of the BH is $ 4.05\,\msun $. The BH begins with a dimensionless pre-merger spin of $a = 0.087$. Here, we describe the neutron star using an SFHo equation of state~\citep{Steiner:2013ApJ...774...17S}. The simulation evolves for just under 5.75 orbits before the system reaches merger. Upon evolving the system to 10 ms post-merger, the BH exhibits a mass of $\mbh = 5.26\,\msun $ and a spin of $ a = 0.59 $. The disrupted NS results in a disk of mass of $ \mdisk = 0.007\,\msun $.

At 10 ms after the collapse, we follow the scheme described in \citet{Gottlieb2023b} to remap the numerical relativity output to the GPU-accelerated GR-MHD code \textsc{h-amr} \citep{Liska2022}, where we simulate the post-merger evolution for an additional $ \sim 1\,\s $.
At the time of remapping, we introduce magnetic fields in the accretion disk. We explore various field configurations, where the geometry can be either toroidal or poloidal, and the field profile depends on the radius and mass density with a cutoff at $ 5\times 10^{-4} $ of the maximum density at the time of remapping. We verify that the fastest-growing MRI mode's wavelength is resolved in all simulations at all times.
Table~\ref{tab:models} summarizes the considered configurations.

\begin{table}[!hp]
		\setlength{\tabcolsep}{25pt}
		\centering
		\renewcommand{\arraystretch}{1.2}
		\begin{tabular}{| c | c c c c c c | c c | }
			
                \hline
			Model & Merger & $ q $ & $ \Mtot $ & $ A $ & $ \beta_p $ & $ t_f\,[\s] $
			\\	\hline
			$ NN-P_w $ & BNS & 1.7 & $ 2.84 $ & $ A_\varphi \propto \rho^2 r^3 $ & $ 10^{3.5} $ & $ 1.3 $ \\ 
			$ NN-P_c $ & BNS & 1.7 & $ 2.84 $& $ A_\varphi \propto \rho^2 r^6 $ & $ 10^{2.5} $ & $ 1.2 $ \\   
			$ NN-P_s $ & BNS & 1.7 & $ 2.84 $& $ A_\varphi \propto \rho^2 r^3 $ & $ 10^{1.5} $ & $ 0.8 $ \\   
			$ NN-T_c $ & BNS & 1.7 & $ 2.84 $& $ A_\theta \propto \rho r^2 $ & $ 10 $ & $ 1.3 $ \\ 
			$ NN-T_s $ & BNS & 1.7 & $ 2.84 $& $ A_\theta \propto \rho r^2 $ & $ 1 $ & $ 1.2 $ \\ 

            $ BN-P_w $ & BH--NS & 3.0 & $ 5.4 $ & $ A_\varphi \propto \rho r^2 $ & $ 10^5 $ & $ 1.0 $ \\ 
			$ BN-P_c $ & BH--NS & 3.0 & $ 5.4 $& $ A_\varphi \propto \rho r^2 $ & $ 10^3 $ & $ 0.7 $ \\   
			$ BN-P_s $ & BH--NS & 3.0 & $ 5.4 $& $ A_\varphi \propto \rho r^2 $ & $ 10^2 $ & $ 1.0 $ \\   
			$ BN-T_c $ & BH--NS & 3.0 & $ 5.4 $& $ A_\theta \propto \rho r^2 $ & $ 10^2 $ & $ 1.2 $ \\ 
			$ BN-T_s $ & BH--NS & 3.0 & $ 5.4 $& $ A_\theta \propto \rho r^2 $ & $ 1 $ & $ 1.1 $ \\ 
                \hline
		\end{tabular}
		
		\caption{
			A summary of the models' parameters. The model names stand for merger type BNS (NN) or BH--NS (BN), and poloidal ($P$) or toroidal ($T$) initial magnetic field, with the subscripts indicating the strength of the field: weak ($w$), canonical ($c$), or strong ($s$). $ q $ is the mass ratio, $ \Mtot $ is the total binary mass, $ A $ is the vector potential, $ \beta_p $ is the characteristic gas to magnetic pressure ratio, and $ t_f $ is the final time of the simulation with respect to the merger time.
    		}
    		\label{tab:models}
	\end{table}

The \textsc{h-amr} grid in spherical-polar coordinates is uniform in $\log r$, $\theta$ and $\varphi$, extending from $ r = r_g $ to $ r = 10^5\,r_g$. The base grid resolution is $N_r\times N_\theta\times N_\varphi = 384\times 96\times 96$ cells. Using static mesh refinement, we double the base resolution (quadruple in model $ T_c $) in all dimensions at $ 4 < r/r_g < 100 $. By using 3 levels of adaptive mesh refinement, we properly resolve the relativistic outflows.

\begin{figure*}
    \centering
    	\includegraphics[width=6.8in]{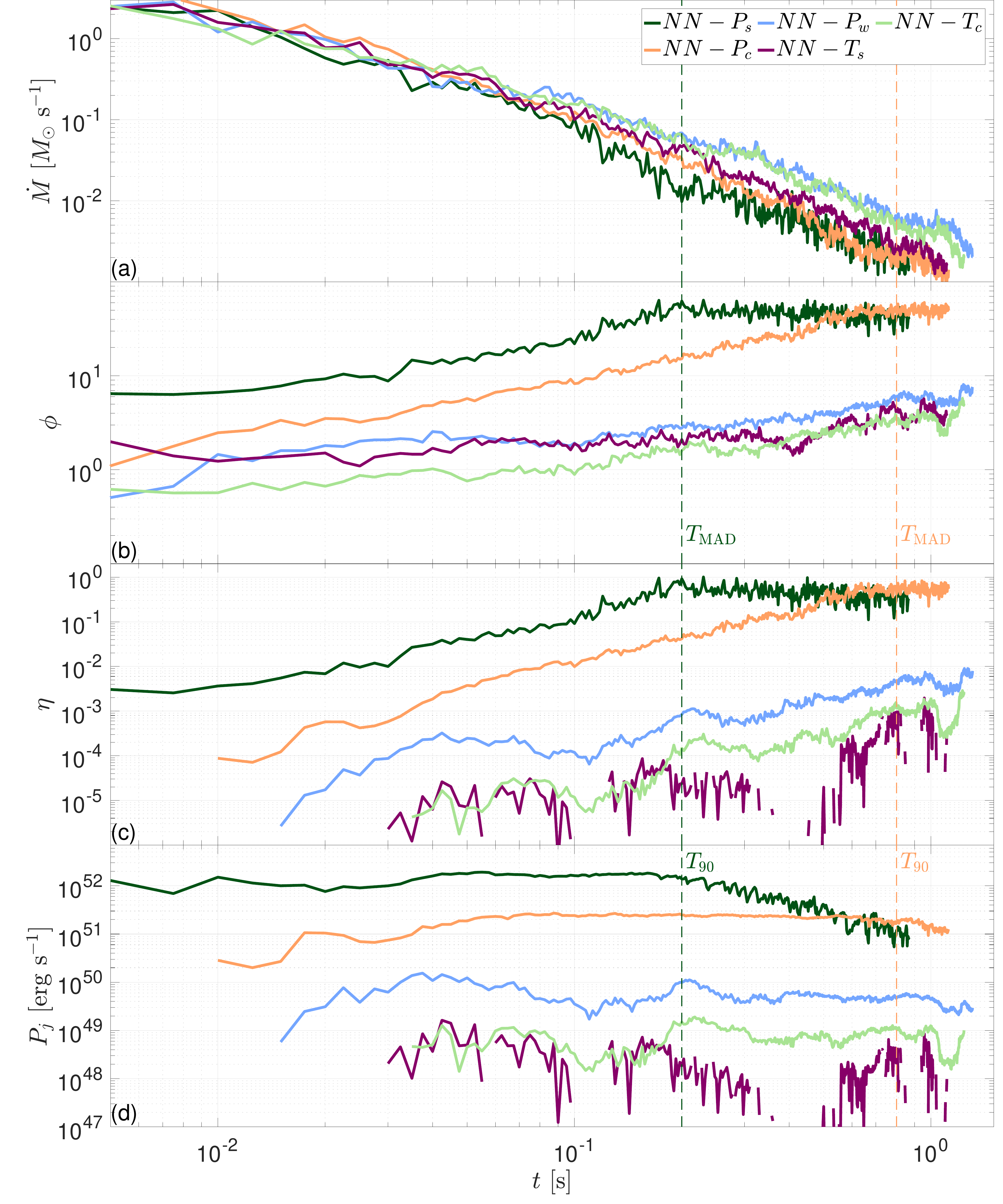}
     \caption{
     Time evolution on the BH horizon for different models.
     {\bf Panel (a)}: The mass accretion rate in all models follows $ \dot{M} \approx (t/0.01\,\s)^{-2} $.
     {\bf Panel (b)}: The dimensionless magnetic flux $ \phi = \Phi/\sqrt{\dot{M}cr_g^2} $ shows a gradual increase until entering the MAD state at $ \phi \approx 50 $ (vertical dashed lines), after which it remains roughly constant.
     {\bf Panel (c)}: The jet launching efficiency $ \eta \equiv \eta_\phi\eta_a $ increases gradually with $ \phi $, until it reaches the maximum launching efficiency $ \eta \approx 0.4 $ in the MAD state.
     {\bf Panel (d)}: The jet power, $ P_j = \eta\dot{M}c^2 $, is roughly constant due to constant dimensional magnetic flux threading the BH. Once the disk becomes MAD, $ \eta $ saturates, and the jet power drops as $ P_j \sim \dot{M} \sim t^{-2} $ (vertical dashed lines).
     }
     \label{fig:BNSsim}
    \end{figure*}

\begin{figure*}
    \centering
    \includegraphics[width=6.8in]{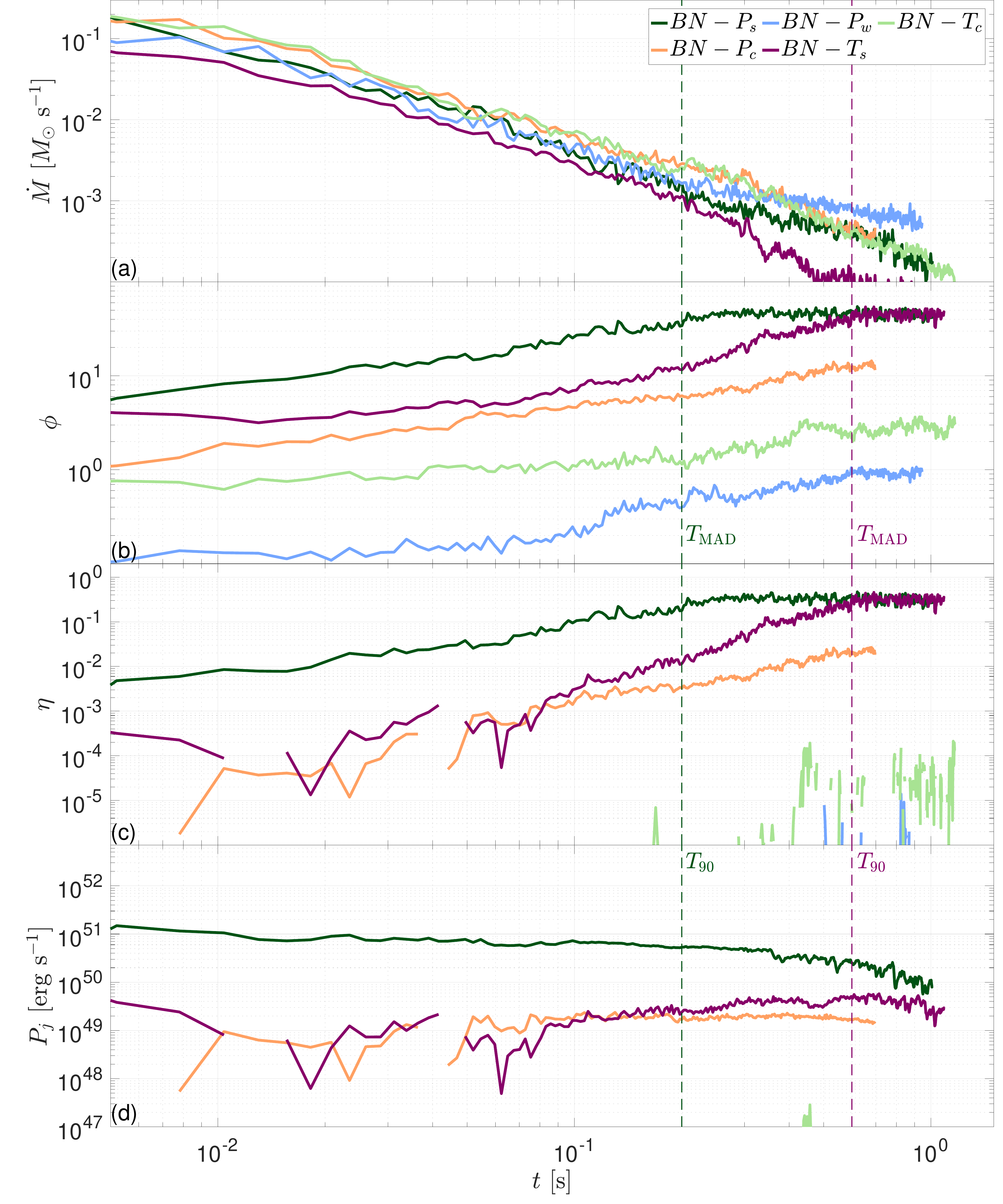}
     \caption{
     Same as Fig.~\ref{fig:BNSsim}, but for the BH-NS merger simulations.
     }
     \label{fig:BHNSsim}
    \end{figure*}

Figures~\ref{fig:BNSsim} and \ref{fig:BHNSsim} depict the temporal evolution of various properties on the BH horizon in the BNS and BH-NS merger simulations, respectively. Panels (a) display the mass accretion rate, featuring a power-law decay of $ \dot{M} \sim t^{-2} $. Panels (b) illustrate that the constant dimensional flux threading the BH leads to a gradual growth of the dimensionless flux $ \phi $ with the decline in $ \dot{M} $. Consequently, the jet launching efficiency, $ \eta = \eta_a\eta_\phi $ steadily increases, as shown in Panels c. Once the flux reaches saturation at $ \phi \approx 50 $ (vertical dashed lines), or $ \eta = \eta_a \approx 0.3~(0.4) $ for $ a = 0.59~(0.68) $, the disk enters a MAD state, and the BH achieves its maximum jet launching efficiency.
As indicated in panels (d), the jet power remains roughly constant at all times before the disk turns MAD, owing to a constant $ \Phi $ on the BH (Eq.~\eqref{eq:BZ}). As demonstrated by the strong poloidal field models, $ P_s $ and $ P_c $, the saturation of $ \phi $ (or drop in $ \Phi $) at $ t > \tmad $ leads to $ P_j \sim t^{-2} $ (Eq.~\eqref{eq:Pj}). Eventually, all models will reach a MAD state within several seconds, marking the typical prompt duration. Models with initial toroidal magnetic configuration in the disk exhibit stronger variability in the light curve, offering a possible origin of the variability observed in \cbGRB light curves.

\end{document}